\documentclass[aps,prb,showpacs,twocolumn]{revtex4}
\usepackage{graphicx}

\input{epsf}

\begin{document}

\title{Photon-assisted spin transport in a two-dimensional electron gas }
\author{M. V. Fistul$^1$ and K. B. Efetov$^{1,2}$}
\affiliation{$^1$ Theoretische Physik III, Ruhr-Universit\"at Bochum, D-44801 Bochum,
Germany}
\affiliation{$^2$ L. D. Landau Institute for Theoretical Physics, 117940 Moscow, Russia}
\date{\today}

\begin{abstract}
We study spin-dependent transport in a two-dimensional electron gas subject
to an external step-like potential $V(x)$ and irradiated by an
electromagnetic field (EF). In the absence of EF the electronic spectrum
splits into spin sub-bands originating from the "Rashba" spin-orbit
coupling. We show that the resonant interaction of propagating electrons
with the component EF parallel to the barrier induces a \textit{%
non-equilibrium dynamic gap} $(2\Delta _{R})$ between the spin sub-bands.
Existence of this gap results in coherent spin-flip processes that lead to a
spin-polarized current and a large magnetoresistance, i.e the spin valve
effect. These effects may be used for controlling spin transport in
semiconducting nanostructures, e.g. spin transistors, spin-blockade devices
etc. , by variation of the intensity $S$ and frequency $\omega $ of the
external radiation.
\end{abstract}

\maketitle


\section{Introduction}

Study of a spin-dependent transport in diverse mesoscopic systems,
e.g. junctions with ferromagnetic layers, magnetic semiconductors,
and low-dimensional semiconducting nanostructures remains one of
the most popular topics during the last decades
\cite{AwschBook,Mres,NatureContr}. Such systems display
fascinating spin-dependent phenomena, e.g. giant magnetoresistance
(the spin valve effect) \cite{Mres,NatureContr}, spin-polarized
current \cite{SPcurr,Fink,MSilv}, and spin-transistor effects
\cite{DD,GrundlerPRB,Loss}, just to name a few. Application of
these effects has resulted in the emerging of new technologies
based on using the electron spin (so-called "spintronics").


An interesting and important example of how the spin-dependent
transport can be realized experimentally is a two-dimensional
electron gas (2DEG) formed in a semiconducting quantum well. In
this particular system a strong spin-orbit interaction is induced
by inhomogeneous electric field in the direction perpendicular to
the 2DEG plane, which is usually referred to as the Rashba effect
\cite{Rashba}. The Rashba spin-orbit interaction results in a
splitting of the 2DEG electronic spectrum $\epsilon (p)$
\begin{equation}
\epsilon _{\pm }(p)=\frac{\mathbf{p}^{2}}{2m}\pm \alpha |\mathbf{p}|,
\label{Spectrum}
\end{equation}%
where $\mathbf{p}=\{p_{x},p_{y}\}$ is the quasi-particle momentum, $m$ is
the effective mass, and $\alpha $ is the strength of the spin-orbit
interaction. The signs $\pm $ in Eq. (\ref{Spectrum}) correspond to
different electron spin projections.

Dynamics of the spins in the presence of the spin-orbit interaction is
characterized by a spin precession around the direction perpendicular to the
momentum $\mathbf{p}$ (in the 2DEG plane). This precession is the basis for
diverse proposals for spin transistors with the use of \textit{homogeneous}
2DEG \cite{DD,GrundlerPRB,Loss}. An additional control of a spin-dependent
transport can be obtained by variation of the spin precession axis. This
goal may be achieved by creating artificially a coordinate dependent
potential $V(x)$ \cite{MSilv}. Such a potential can be produced by using a
split-gate technique or cleaved edge fabrication method \cite{Techn}.

In the static case, as no time-dependent fields are applied, the
precession frequency and the corresponding splitting between the
spin sub-bands are determined by both the transverse quantization
of the momentum $\mathbf{p}$, and a total change of the potential
$V\left( x\right) $ \cite{MSilv}. Such a setup might allow one to
produce the spin-polarized current for quasi-particles with
nonzero values of the transverse momentum. However, to observe
this effect a small value of $\epsilon _{0}-V~\simeq ~m\alpha
^{2}$  proportional to the small parameter $\alpha ^{2}$ has to be
used ($\epsilon _{0}$ is the Fermi energy) \cite{MSilv}. Moreover,
the quasi-particles propagating in the direction perpendicular to
the barrier are not spin-polarized.

In this paper we suggest a new method of producing spin-polarized
current using a similar structure with the 2DEG and a step-like
potential $V\left( x\right) $. However, in addition to the
previous set up, we assume that the system is irradiated by an
external electromagnetic field (EF). We demonstrate that in this
situation the ballistic transport of the quasi-particles moving
perpendicular to the barrier can be extremely sensitive to the
time-dependent perturbation provided certain resonant conditions
are met. Notice here that although our analysis is applied
directly to the Rashba type of a spin-orbit interaction, similar
effects can be obtained also in systems with a bulk asymmetry
displaying Dresselhaus type of a spin-orbit interaction
\cite{Dresselhaus}

\section{Model and qualitative analysis}

To be specific, we consider 2DEG subject to an external potential
$V(x)$, and in the presence of an EF applied in the transverse
direction parallel to the potential barrier. The system is
represented in Fig. 1. We stress here that, although the EF need
not be linearly polarized, only the component of EF parallel to
the interface ($y$-direction) leads to the resonant interaction
between the spin sub-bands.

The resonance condition can be written as $\hbar \omega =2\alpha |\mathbf{p}%
(x)|$, where $\omega $ is the frequency of EF, and $\mathbf{p}(x)$ is the
coordinate dependent classical momentum of the quasi-particles. Such a
resonant interaction leads to forming a \textit{non-equilibrium dynamic gap}
$(2\Delta _{R})$ between the spin sub-bands. The quantity $\Delta _{R}/\hbar
$ has the same meaning as the famous Rabi frequency for microwave induced
quantum coherent oscillations between two energy levels \cite{Hanggi} (these
energy levels are $\frac{|\mathbf{p}\left( x\right) |^{2}}{2m}+\alpha |%
\mathbf{p}\left( x\right) |$ and $\frac{|\mathbf{p}\left( x\right) |^{2}}{2m}%
-\alpha |\mathbf{p}\left( x\right) |$ in our case). The value of the gap
depends strongly on the intensity $S$ and frequency $\omega $ of the
external radiation.

The dynamic gap induces coherent spin-flip processes and manifests
itself in generating a spin polarization of the current. This may
also lead to a strong suppression of the conductivity $G$ of 2DEG
with spin-polarized (ferromagnetic) leads, i.e. the spin-valve
effect (see Fig. 1).

\begin{figure}[tbp]
\includegraphics[width=2.7in,angle=-90]{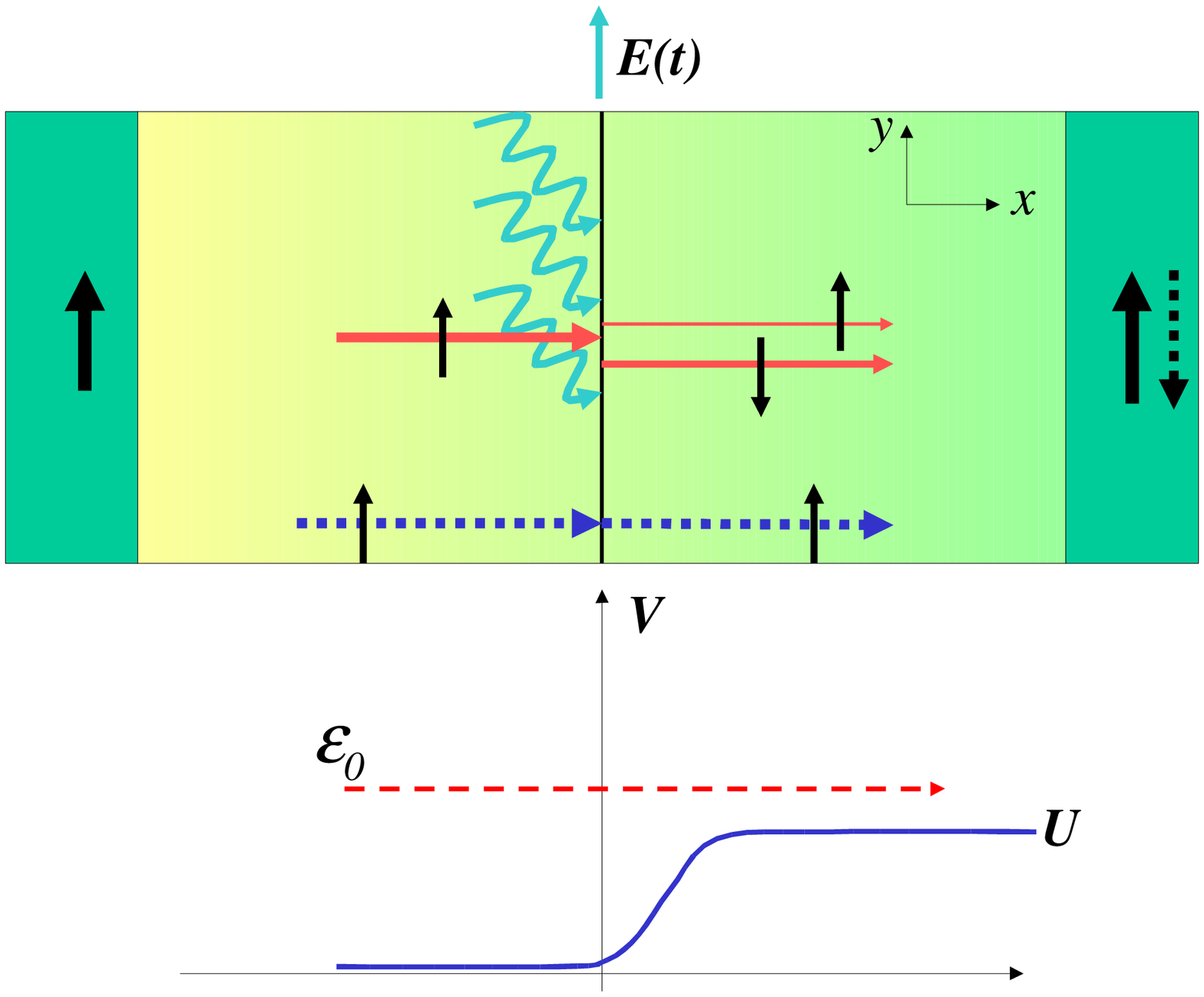} %
\caption{(color online) Transport in a 2D electron gas with
spin-polarized electrodes. The ferromagnetic (anti-ferromagnetic)
configuration of the leads is shown. The 2D electron gas interacts
with an external potential $V(x)$ and is irradiated by
electromagnetic field (EF). The dashed (solid) lines display spin
transfer in the absence (presence) of EF.} \label{Schematic}
\end{figure}

\section{Floquet eigenvalues of the problem}
We start our analysis writing a time and coordinate dependent two spin-band
Hamiltonian $\hat{H}\left( t\right) $ in the external EF
\begin{equation}
\hat{H}(t)=\frac{\left[\mathbf{\hat{p}}-\frac{e}{c}\mathbf{A}\right]^2}{2m}+\alpha \left[ \mathbf{\hat{\sigma}%
}\times \left\{ \mathbf{\hat{p}-}\frac{e}{c}\mathbf{A}\left(
t\right) \right\} \right] _{z}+V(x),  \label{Hamiltonian}
\end{equation}%
where the $\mathbf{\hat{\sigma}}=\{\hat{\sigma}_{x},\hat{\sigma}_{y}\}$ are
the standard Pauli matrices. The electromagnetic wave is represented by the $%
y$-component of the vector-potential as $A_{y}=(Ec/\omega )\cos (\omega t)$,
where $E=\sqrt{4\pi S/c}$ is the amplitude of the electric field.

Next, we take into account the {\it resonant interaction} between
the EF and propagating quasi-particles, only. Therefore, we
neglect a weak non-resonant interaction of quasi-particles with EF
that is e.g. due to the presence of a time-dependent
vector-potential $\mathbf{A}$ in the first term of the Hamiltonian
(\ref{Hamiltonian}). In this case the time-dependent problem described by the Hamiltonian (\ref%
{Hamiltonian}) is reduced to a stationary problem by switching to
a rotating frame with the following unitary transformation of the
two component wave functions
\begin{equation}
\hat{U}_{n}=\frac{1}{\sqrt{2}}\left(
\begin{tabular}{cc}
$1$ & \hspace{0.2cm} $1$ \\
$-\exp (-i\hat{\theta})$ & \hspace{0.2cm} $\exp (-i\hat{\theta})$%
\end{tabular}%
\ \right) \exp \left[ i\omega t\left( n-\frac{\hat{\sigma}_{z}+1}{2}\right) %
\right] ,  \label{a2}
\end{equation}%
where $\hat{\theta}=tan^{-1}(\hat{p}_{x}/\hat{p}_{y})$ and $\hat{p}_{x},$ $%
\hat{p}_{y}$ are the components of momentum operators
perpendicular and parallel to the interface, respectively. A
similar procedure has been used
to analyze the transport in a graphene layer in the presence of EF \cite%
{graphene} but spin degrees of freedom were irrelevant in that consideration.

The transformation, Eq. (\ref{a2}), changes the initial Hamiltonian to $\hat{%
H}_{eff}^{\prime }=\hat{U}_{n}^{+}H\hat{U}_{n}-i\hbar \hat{U}_{n}^{+}\hat{%
\dot{U}}_{n}$. The latter contains, in general, both static and proportional
to $\exp \left( \pm 2i\omega t\right) $ parts. However, like for the two
level systems \cite{Hanggi}, only the static part of $\hat{H}_{eff}$ is
important near the resonance, and it can be written as%
\[
\hat{H}_{eff}=\frac{|\mathbf{\hat{p}}|^{2}}{2m}+V(x)+
\]%
\begin{equation}
+\left(
\begin{tabular}{cc}
$\hbar (n-1)\omega +\alpha |\mathbf{\hat{p}}|$ & \hspace{0cm} $\frac{%
eE\alpha }{2\omega }$ \nonumber\vspace{0.2cm} \\
$\frac{eE\alpha }{2\omega }$ & \hspace{0cm} $\hbar n\omega -\alpha |\mathbf{%
\hat{p}}|$%
\end{tabular}%
\ \right) ~,  \label{a3}
\end{equation}%
where $|\mathbf{\hat{p}}|=\sqrt{\hat{p}_{x}^{2}+\hat{p}_{y}^{2}}$.

Neglecting the oscillating part of the Hamiltonian
$\hat{H}_{eff}^{\prime }$ corresponds to a rotation wave
approximation (RWA) \cite{Hanggi}. The RWA is valid in the most
interesting regime of the resonant interaction between the EF and
propagating quasi-particles when
\begin{equation}
\hbar \omega \simeq 2\alpha |\mathbf{p}\left( x\right) |  \label{a100}
\end{equation}%
We also assume that the amplitude of the external
microwave radiation is comparatively small, $eE\alpha /\hbar \ll
\omega ^{2}$.

The Eq. (\ref{a3}) shows that the EF results in the appearance of
off-diagonal elements in the operator $\hat{H}_{eff}$. In the absence of the
coordinate dependent potential, i.e. $V(x)=0$, the eigenvalues $\tilde{%
\epsilon}(p)$ of $\hat{H}_{eff}$ give the sets of bands of quasi-energies
(the Floquet eigenvalues \cite{Hanggi}):
\begin{equation}
\tilde{\epsilon}_{n,\pm }(p)~=~\frac{\mathbf{p}^{2}\left( x\right) }{2m}+(n-%
\frac{1}{2})\hbar \omega \pm \sqrt{(\alpha |\mathbf{p}\left( x\right) |-%
\frac{\hbar \omega }{2})^{2}+\Delta _{R}^{2}}  \label{Eigenvalues}
\end{equation}%
where
\begin{equation}
2\Delta _{R}=(e\alpha /\omega )\sqrt{4\pi S/c}  \label{a4}
\end{equation}%
is the EF induced non-equilibrium gap and $n$ are integer numbers
$n=0,\pm 1,\pm 2,...$. It is well known \cite{Hanggi} that, in the
presence of periodic time-dependent perturbations, the bands of
the Floquet eigenvalues replace the quasi-particle spectrum, Eq.
(\ref{Spectrum}). The typical bands of Floquet eigenvalues are
shown in Fig. 2.
\begin{figure}[tbp]
\includegraphics[width=3in]{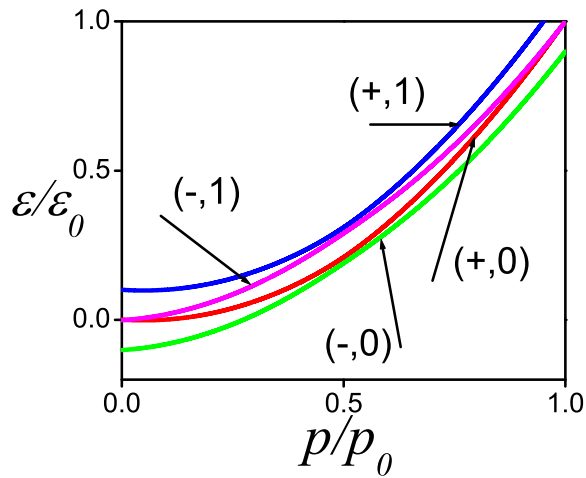}
\caption{(color online) The bands of Floquet eigenvalues for $n=0,
1$ are shown. Each band is characterized by a sign in the Eq.
(\ref{Eigenvalues}) and integer number $n$. The parameters $\hbar
\omega=\alpha p_0=0.1\epsilon_0$, $\Delta_R=0.01 \epsilon_0 $ were
chosen. The momentum $p_0$ is defined as $\epsilon_0=p_0^2/(2m)$.
} \label{EiFl}
\end{figure}

\section{photon assisted spin dynamics: quasi-classical description }
Next, we analyze the spin-dependent transmission of quasi-particles $%
P_{\uparrow (\downarrow ),\uparrow (\downarrow )}$ through the
potential barrier $V(x)$ formed in the 2DEG. Here, the $\uparrow$
($\downarrow$) corresponds to positive (negative) values of a spin
projection on the $y$-axis. To obtain the analytical solution we
use the quasi-classical approximation that can be quite realistic
in the presence of a smooth potential created electrostatically.
The classical phase trajectories $p(x)$ of the Hamiltonian
$\hat{H}_{eff}$ are determined by the conservation of the sum of
the potential energy $V(x)$ and the quasi-energy
$\tilde{\epsilon}_{n,\pm }(p)$ as
\begin{equation}
V(x)+\tilde{\epsilon}_{n,\pm }(p)=\epsilon _{0}~~  \label{energyconserv}
\end{equation}%
Using Eqs. (\ref{Eigenvalues}) and (\ref{energyconserv}) for
$n=0,1$ we obtain four spin-dependent phase trajectories
characterized by a sign in the Eq. (\ref{Eigenvalues}) and integer
number $n$.  A further progress can be made by choosing a specific
model for the electrostatic potential formed in 2DEG ($d$ is the
characteristic width of the potential) \cite{Tunn}
\begin{equation}
V(x)~=~\left\{
\begin{array}{cc}
0, & x<0 \\
Fx, & 0<x<d \\
U=Fd, & x>d%
\end{array}%
\right.  \label{Potential}
\end{equation}%
Since the influence of EF is diminished for quasi-particles
possessing a large transverse momentum $p_y~\simeq~p_x$, we just
consider a one-dimensional transport with $p_x \gg p_y$.
The typical phase trajectories for the potential $V\left( x\right) $, Eq. (%
\ref{Potential}), are shown in Fig. 3.
\begin{figure}[tbp]
\includegraphics[width=3in,angle=0]{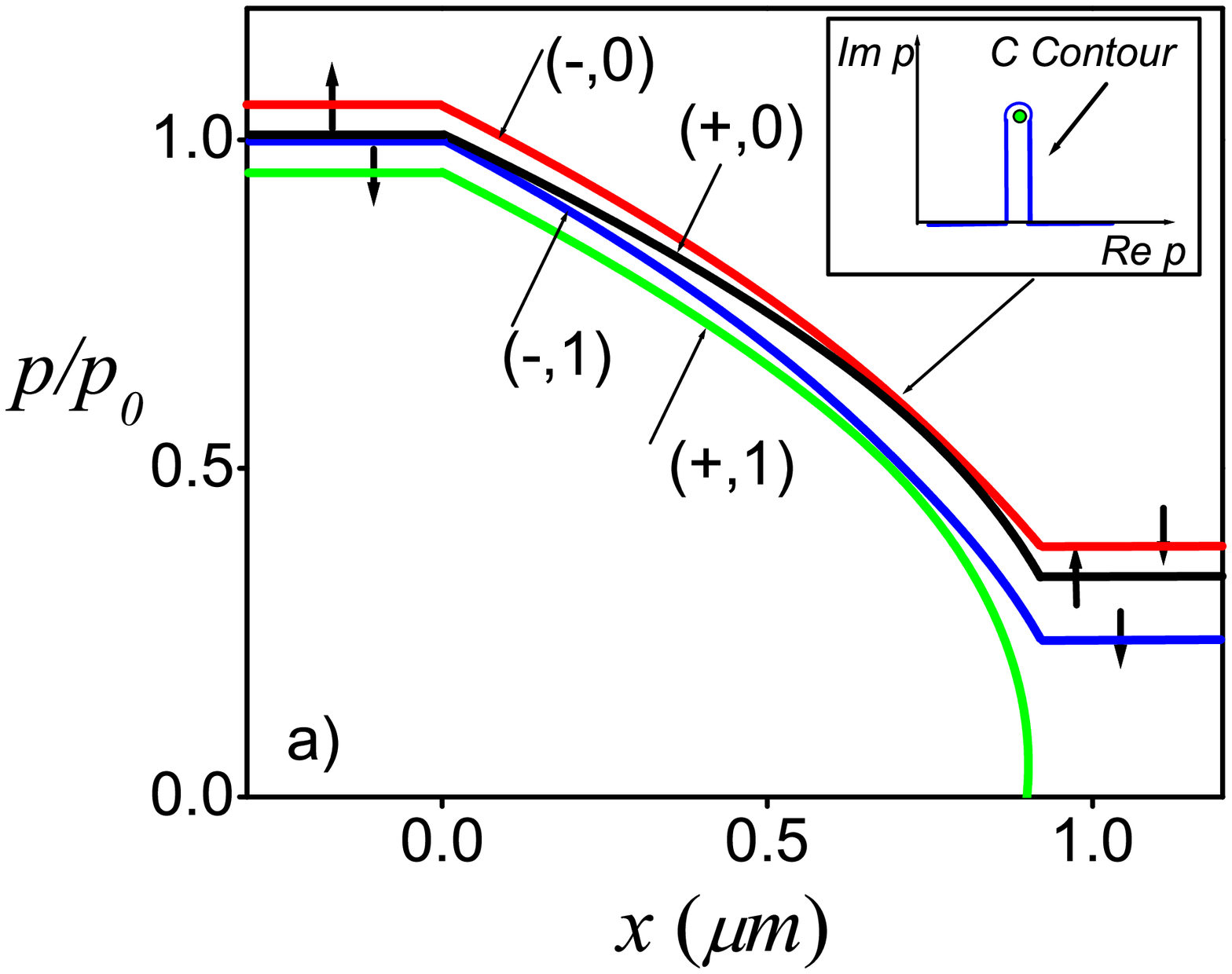}
\end{figure}
\begin{figure}[tbp]
\includegraphics[width=3in]{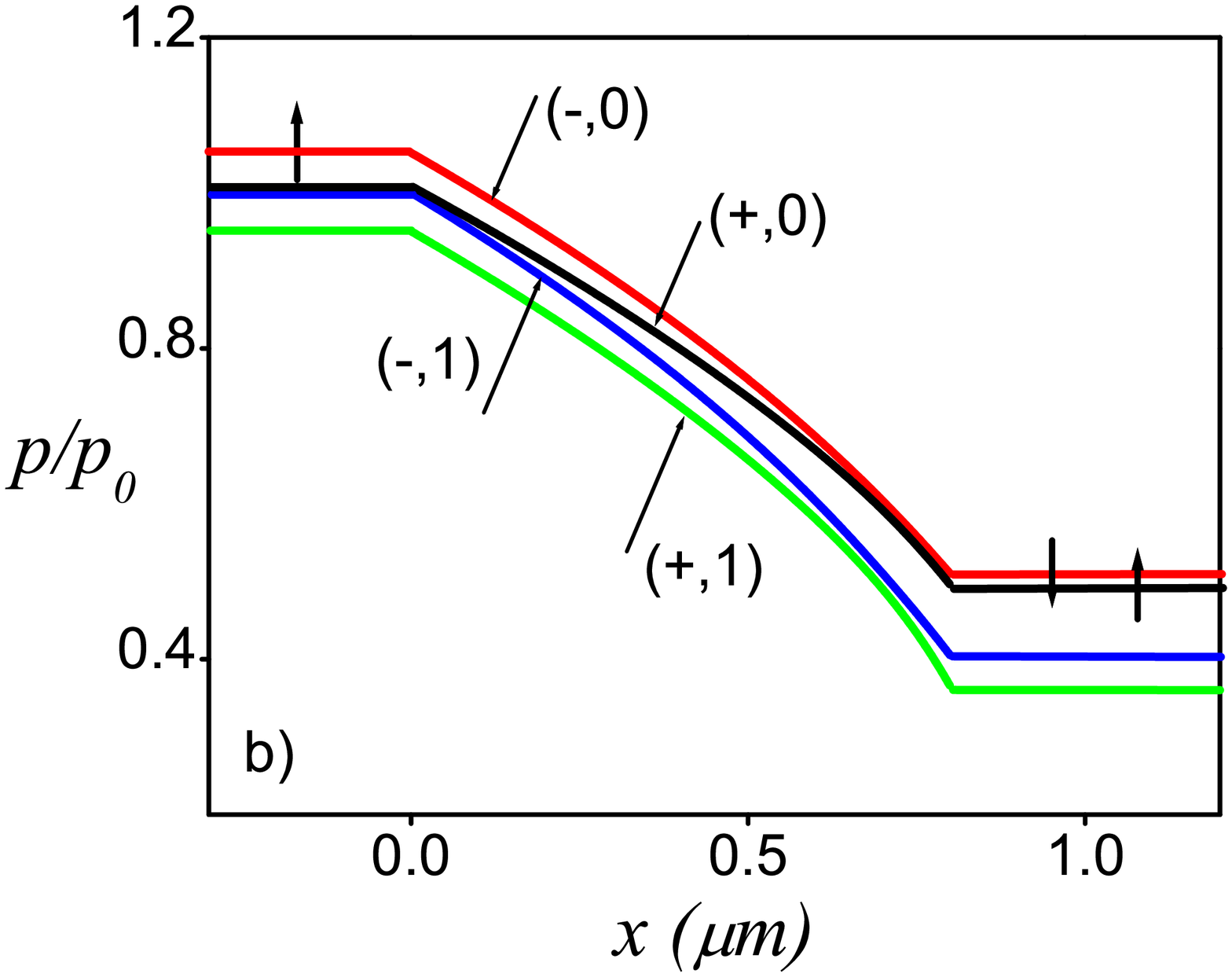}
\caption{(color online) Spin-dependent phase trajectories $p(x)$ of the Hamiltonian $\hat{H}%
_{eff}$ for two cases: a) spin-polarized current is established,
the potential barrier height $U=0.92\protect\epsilon _{0}$ was
used; b) the
magnetoresistance effect is realized, the potential barrier height $U=0.8%
\protect\epsilon _{0}$ was used. Insert shows the contour of integration in
the complex plane used when calculating the "tunneling" process between
adjacent phase trajectories. The values of parameters $\hbar \protect%
\omega =\alpha p_0=0.1\protect\epsilon _{0}$, $d=1\protect\mu m$ and $\Delta _{R}=0.01%
\protect\epsilon _{0}$ were chosen.}
\label{PhaseTr}
\end{figure}

As the quasi-particles approach the barrier, the momentum $p_{x}$
decreases, the resonance condition is satisfied sufficiently close
to the junction, and, therefore, the interaction with the EF opens
the gap $2\Delta _{R}$ between neighboring trajectories. In this
case the propagation along the phase trajectories corresponds to
the \textit{coherent spin-flip processes} , and the spin
conservation is possible only due to the switching ("tunneling")
between adjacent trajectories (see Fig. 3). In the regime of
strong induced spin-flip processes, the probability of such
dynamical tunneling $P_{tun}$ is small and $P_{tun}$ is obtained
by a shift of the integration contour $C$ in the complex $p$ plane
around the branch point ($\tilde{p}=\hbar \omega /(2\alpha
)+i\Delta _{R}/v\alpha $) \cite{graphene,Tunn,Kadig} (see, insert
in Fig. 3a) as
\begin{equation}
P_{tun}=|\exp \left\{ i\frac{2}{\hbar }\int p(x)dx\right\} |=|\exp \left\{ i%
\frac{2}{\hbar F}\int_{C}\tilde{\epsilon}_{n}(p)dp\right\} |,  \label{Newint}
\end{equation}%
%
%
%
%
%

Substituting the expressions for the Floquet eigenvalues, Eq. (\ref%
{Eigenvalues}), into Eq.(\ref{Newint}), and calculating the integral in Eq. (%
\ref{Newint}) we write the probability of the dynamical tunneling $P_{tun}$
as
\begin{equation}
P_{tun}\simeq \exp \left[ -\frac{\pi \Delta _{R}^{2}}{\hbar \alpha F}\right]
~,  \label{Tunnsupp}
\end{equation}%
%
%
where the gap $2\Delta _{R}$ should be taken from Eq. (\ref{a4}).

The probability of the spin-flips $P_{sf}$ is given by the following
expression $P_{sf}=1-P_{tun}$. Therefore, the external radiation of the
frequency $\omega \sim ~\alpha p_{0}/\hbar $, where $p_{0}$ is the Fermi
momentum, satisfying the resonant condition can induce strong spin-flip
processes as $P_{sf}\simeq 1$. In the opposite case of a large frequency, $%
\omega \geq 2\alpha p_{0}/\hbar $, the EF cannot provide the resonant
interaction, and the propagation of quasi-particles moving perpendicular to
the barrier is not spin-dependent.

\section{Discussion and conclusions}
Using Eqs. (\ref{energyconserv}, \ref{Newint}) we see that a
spin-polarized current can be created provided potential step $U$
is sufficiently high, such that $\epsilon _{0}-U<\hbar \omega $ .
Indeed, as it is shown in Fig. 3a, the quasi-particles moving
along the lower phase trajectory, are reflected from the barrier.
Therefore, the probability $P_{\downarrow ,\uparrow }=0$. Other
spin-dependent probabilities of quasi-particles propagation are
$P_{\uparrow ,\uparrow }=P_{\downarrow ,\downarrow }=P_{tun}$ that
corresponds to the switching between $(-,0) \rightarrow (+,0) $,
and $(+,1) \rightarrow (-,1)$ trajectories (see Fig. 2a). The
quasi-particle motion along the upper phase trajectory is
characterized by a probability
 $P_{\uparrow ,\downarrow }=1-P_{tun}$. Therefore, if on the left side of
2DEG the non spin-polarized quasi-particles are induced, the total
polarization of transmitted quasi-particles moving perpendicular
to the barrier equals
\begin{equation}
|<\sigma _{y}>|=|-(1-P_{tun})+P_{tun}-P_{tun}|=1-P_{tun}
\label{Totalpol}
\end{equation}

In experiments, the EF induced spin-flips processes lead also to a large
magnetoresistance effect in the conductivity $G$ of a 2DEG with
spin-polarized electrodes (see Fig. 1). We use the simplest setup with the
magnetization in the leads directed along the $y$ axis. In this case, we can
neglect the quantum-mechanical interference between the different spin
states, and the conductivity $G$ is determined by the probability of
spin-flip processes, $P_{sf}$. Moreover, the conductivity depends on the
relative directions of magnetization axes in the leads. Thus, in the case of
the ferromagnetic configuration of the leads , i.e when the directions of
the magnetization in both the electrodes coincide, the conductivity $%
G_{\uparrow ,\uparrow }~\propto ~P_{tun}$ is strongly suppressed.
On the contrary, in the case of the anti-ferromagnetic
configuration of the leads, i.e. when the directions of
magnetization in electrodes are opposite to each other, the
conductivity $G_{\uparrow ,\downarrow }~\propto ~(1-P_{tun})$ is
not suppressed. The diverse phase trajectories characterizing the
spin dynamics for such a case are shown in Fig. 3b. These results
are valid provided the potential barrier is not too high, i.e
$\hbar \omega <\epsilon _{0}-U<\epsilon _{0}(\frac{\hbar \omega
}{2\alpha p_{0}})^{2}$. Notice here, that in contrast to diverse
spin valve devices \cite{Mres,NatureContr} the photon-assisted
magnetoresistance displays negative value.

Finally, we address the question of experimental conditions
necessary to observe the predicted effects. An experimental setup
has to provide a ballistic transport in a 2DEG in order to avoid
the incoherent spin-flip processes. In systems with spin-orbit
coupling these processes arise due to the presence of randomly
distributed impurities (the Dyakonov-Perel spin flip mechanism
\cite{Dyak}). Therefore, the elastic scattering of quasi-particles
diminishes the obtained effects. Moreover, the elastic scattering
and/or all mechanisms leading to the inhomogeneous broadening of
the EF can also directly modify the probability of photon-assisted
spin-flip processes i.e. the $P_{tun}$. This effect demands a
separate analysis and goes beyond a scope of this paper. A barrier
with the typical width $d~\simeq ~1\mu m$ in a 2DEG has to be
fabricated. An external radiation containing the component
parallel to the barrier of a moderate intensity $S$ has to be
applied. We emphasize that EF need not be linearly polarized. The
EF induces the spin-flips processes and corresponding
spin-dependent quasi-particle
transmission through the barrier in the range of the frequencies of EF $%
\omega ~\simeq ~\alpha p_{0}$. For example in the $InAs$-based
heterostructures characterized by a strong spin-orbit coupling \cite%
{Grundler}, the spin-dependent transport can be controlled by the
EF with the frequency $\simeq ~1~THz$. To observe intensive
resonantly induced spin-flip processes, and therefore, a
spin-dependent transport one may use the radiation with a moderate
intensity $S~>0.4~W/cm^{2}$. In order to obtain a spin-polarized
current the barrier has to be rather high: $\epsilon _{0}-U<\hbar
\omega ~\simeq ~3meV$. The magnetoresistance effect can be
observed for moderate values of the barrier: $3meV<\epsilon
_{0}-U<\epsilon _{0}/4~\simeq ~10meV$.

In conclusion, we have demonstrated that the resonant radiation of
a moderate intensity $S$ applied to a ballistic 2DEG with the
spin-orbit interaction leads to a spin-polarized current and/or a
strong magnetoresistance effect in the conductivity of a 2DEG with
ferromagnetic leads. These effects occur due to formation of a
non-equilibrium dynamic gap between spin-subbands in the
quasi-particle spectrum as the resonant condition $\hbar \omega
=2\alpha |\mathbf{p}|$ is satisfied. In the presence of spatially
dependent potential formed in a 2DEG, such a resonance condition
is satisfied somewhere in the vicinity of the barrier. The
presence of the gap induces intensive coherent spin-flips
processes on the barrier. In order to conserve the spin value the
quasi-particles have to "tunnel" between the quasi-classical
trajectories. The value of the gap and, therefore, this specific
type of the tunnelling is controlled by variation of the intensity
$S$ and frequency $\omega $ of the external radiation. We hope
that the predicted effect will find its applications to
non-equilibrium (photon-assisted) spintronics devices based on a
2DEG.

We would like to thank P. Silvestrov , A. Kadigrobov and S.
Syzranov for useful discussions and acknowledge the financial
support by SFB 491.

\end{document}